\documentclass[
reprint,
superscriptaddress,
amsmath,
amssymb,
aps,
prl,
floatfix,
]{revtex4-2}

\usepackage{graphicx}
\usepackage{dcolumn}
\usepackage{bm}
\usepackage[hypertexnames=false]{hyperref}
\hypersetup{
	colorlinks = true,
	allcolors = blue,
}

\usepackage{MnSymbol}
\usepackage[utf8]{inputenc}
\usepackage{multirow}
\usepackage{siunitx}
\usepackage{amsmath, amssymb}
\usepackage{amsfonts}
\usepackage{physics}

\usepackage{bbold}
\usepackage{tikz}
\usetikzlibrary{decorations.pathreplacing}
\usetikzlibrary{backgrounds}
\usetikzlibrary{arrows.meta}
\usetikzlibrary{calc}
\usetikzlibrary{external}
\usepackage{MnSymbol}
\usepackage{float}
\usepackage{adjustbox}
\usepackage[normalem]{ulem}
\usepackage{amssymb}

\usepackage{pgfplots}
\usepgfplotslibrary{colorbrewer}
\usepgfplotslibrary{groupplots}
\pgfplotsset{
	small,
	compat=newest
}
\usepgfplotslibrary{fillbetween}

\newcommand{\Mi}{\mathrm{i}} 
\newcommand{\Me}{\mathrm{e}} 
\newcommand{\Mbf}[1]{\boldsymbol{#1}} 
\newcommand{\Mdiff}{\mathrm{d}} 
 \renewcommand{\vec}[1]{\Mbf{#1}}

\newcommand{\mapsfrom}{\mathrel{\reflectbox{$\mapsto$}}}

\makeatletter
\newcommand{\Distance}[3]{
	\tikz@scan@one@point\pgfutil@firstofone($#1-#2$)\relax  
	\pgfmathsetmacro{#3}{veclen(\the\pgf@x,\the\pgf@y)}
}
\makeatother

\makeatletter
\newcommand{\DistanceCM}[3]{
	\tikz@scan@one@point\pgfutil@firstofone($#1-#2$)\relax  
	\pgfmathsetmacro{#3}{round(0.99626*veclen(\the\pgf@x,\the\pgf@y)/0.0283465)/1000}
}
\makeatother

\tikzset{
	/pgf/images/external info,
	png export/.style={
		external/system call/.add={}{
			&& magick convert -density 1000 -transparent white "\image.pdf" "\image.png"
		},
	},
}

\tikzset{
	/pgf/images/external info,
	use png/.style={png export,png import},
	png import/.code={
		\tikzset{
			/pgf/images/include external/.code={
				\includegraphics
				[width=\pgfexternalwidth,height=\pgfexternalheight]
				{{##1}.png}
			}
		}
	}
}

\makeatletter
\def\pgfplots@getautoplotspec into#1{
	\begingroup
	\let#1=\pgfutil@empty
	\pgfkeysgetvalue{/pgfplots/cycle multi list/@dim}\pgfplots@cycle@dim
	\let\pgfplots@listindex=\pgfplots@numplots
	\pgfkeysgetvalue{/pgfplots/cycle list set}\pgfplots@listindex@set
	\ifx\pgfplots@listindex@set\pgfutil@empty
	\else 
	\c@pgf@counta=\pgfplots@listindex
	\c@pgf@countb=\pgfplots@listindex@set
	\advance\c@pgf@countb by -\c@pgf@counta
	\globaldefs=1\relax
	\edef\setshift{
		\noexpand\pgfkeys{
			/pgfplots/cycle list shift=\the\c@pgf@countb,
			/pgfplots/cycle list set=
		}
	}
	\setshift
	\fi  
	\pgfkeysgetvalue{/pgfplots/cycle list shift}\pgfplots@listindex@shift
	\ifx\pgfplots@listindex@shift\pgfutil@empty
	\else
	\c@pgf@counta=\pgfplots@listindex\relax
	\advance\c@pgf@counta by\pgfplots@listindex@shift\relax
	\ifnum\c@pgf@counta<0
	\c@pgf@counta=-\c@pgf@counta
	\fi
	\edef\pgfplots@listindex{\the\c@pgf@counta}
	\fi
	\ifnum\pgfplots@cycle@dim>0
	\c@pgf@counta=\pgfplots@cycle@dim\relax
	\c@pgf@countb=\pgfplots@listindex\relax
	\advance\c@pgf@counta by-1
	\pgfplotsloop{
		\ifnum\c@pgf@counta<0
		\pgfplotsloopcontinuefalse
		\else
		\pgfplotsloopcontinuetrue
		\fi
	}{
		\pgfkeysgetvalue{/pgfplots/cycle multi list/@N\the\c@pgf@counta}\pgfplots@cycle@N
		\pgfplotsmathmodint{\c@pgf@countb}{\pgfplots@cycle@N}
		\divide\c@pgf@countb by \pgfplots@cycle@N\relax
		\expandafter\pgfplots@getautoplotspec@
		\csname pgfp@cyclist@/pgfplots/cycle multi list/@list\the\c@pgf@counta @\endcsname
		{\pgfplots@cycle@N}
		{\pgfmathresult}
		\t@pgfplots@toka=\expandafter{#1,}
		\t@pgfplots@tokb=\expandafter{\pgfplotsretval}
		\edef#1{\the\t@pgfplots@toka\the\t@pgfplots@tokb}
		\advance\c@pgf@counta by-1
	}
	\else
	\pgfplotslistsize\autoplotspeclist\to\c@pgf@countd
	\pgfplots@getautoplotspec@{\autoplotspeclist}{\c@pgf@countd}{\pgfplots@listindex}
	\let#1=\pgfplotsretval
	\fi
	\pgfmath@smuggleone#1
	\endgroup
}

\pgfplotsset{
cycle list set/.initial=
}
\makeatother

\begin{document}

\title{Conformal duality of the nonlinear Schrödinger equation: \\ Theory and applications to parameter estimation}

\author{David B. Reinhardt}
\email[Corresponding authors email: ]{david.reinhardt@dlr.de; matthias.meister@dlr.de}

\affiliation{German Aerospace Center (DLR), Institute of Quantum Technologies, Wilhelm-Runge-Straße 10, 89081 Ulm, Germany}

\author{Dean Lee}

\affiliation{Facility for Rare Isotope Beams and Department of Physics and
Astronomy,
Michigan State University, MI 48824, USA}

\author{Wolfgang P. Schleich}

\affiliation{Institut f\"ur Quantenphysik and Center for Integrated Quantum Science and Technology (IQST), Universit\"at Ulm, D-89069 Ulm, Germany}

\affiliation{Hagler Institute for Advanced Study at Texas A$\&$M University, Texas A$\&$M AgriLife Research, Institute for Quantum Science and Engineering (IQSE), and Department of Physics and Astronomy, Texas A$\&$M University, College Station, Texas 77843-4242, USA}

\author{Matthias Meister}

\email[Corresponding authors email: ]{david.reinhardt@dlr.de; matthias.meister@dlr.de}

\affiliation{German Aerospace Center (DLR), Institute of Quantum Technologies, Wilhelm-Runge-Straße 10, 89081 Ulm, Germany}

\date{July 4, 2024}

\begin{abstract}
The nonlinear Schrödinger equation (NLSE) is a rich and versatile model, which in one spatial dimension has stationary solutions similar to those of the linear Schrödinger equation as well as more exotic solutions such as solitary waves and quantum droplets. Here we present the unified theory of the NLSE, showing that all stationary solutions of 
the local one-dimensional cubic-quintic NLSE can be classified according to a single number 
called the cross-ratio. Any two solutions with the same cross-ratio can be converted into one another using a conformal transformation, and the same also holds true for traveling wave solutions. Further, we introduce an optimization afterburner that relies on this conformal symmetry to substantially improve NLSE parameter estimation from noisy empirical data.
The new method therefore should have far reaching practical applications for nonlinear physical systems.

\end{abstract}

\maketitle

\textit{Introduction --}
The nonlinear Schrödinger equation (NLSE) is ubiquitous in physics, where it plays a key role in plasma physics~\cite{Bohm_1949a,Bohm_1949b,Hasegawa_1975}, hydrodynamics~\cite{zakharov1968stability,Peregrine_1983,Kuznetsov_1986}, degenerate quantum gases~\cite{Dalfovo1999,Giorgini_2008} and light propagation in nonlinear fiber optics~\cite{Haus_1996,Kivshar_1998,Lederer_2008,copie2020physics}.  
Understanding the possible solutions of the NLSE and estimating its experimental parameters reliably~\cite{motulsky1987fitting, draper1998applied, press2007numerical, raissi2019physics, jiang2022physics} is therefore of great importance for a large variety of purposes whether they are application-oriented or fundamental. 
In this Letter we point out a conformal duality between different classes of solutions and even different orders of the NLSE. 
This conformal mapping provides a unified picture of the cubic- and the cubic-quintic NLSE and even establishes a direct link to the linear Schrödinger equation. 
In this way, our method provides a systematic classification of the complete solution spaces of these equations.
Moreover, the conformal duality can be applied to substantially improve NLSE parameter estimation from noisy experimental data.

The linear Schrödinger equation typically features oscillating and constant-amplitude solutions which have their counterparts in the NLSE. 
However, there also exist solutions which are uniquely nonlinear such as solitary waves~\cite{Zhakharov1971, muryshev2002dynamics,Konotop_2004, petrov2016ultradilute,zhou2021stability,seidel2022conservative} which are of broad interest in physics~\cite{burger1999dark,denschlag2000generating, Strecker_2002,Becker_2008,kibler2015superregular}. Considering (multiple) higher-order self-modulating terms like in the cubic-quintic NLSE drastically expands the solution space allowing for instance for bright and dark soliton pairs~\cite{crosta2011bistability}, droplet solutions~\cite{bulgac2002dilute} and solitons with power law tail decay \cite{hayata1995algebraic}. 
Although the different polynomial NLSEs have been studied in great detail~\cite{akhmediev1987exact, gagnon1989exact,Pushkarov_1996,schurmann1996traveling,Serkin_2000,carr2000stationary1,carr2000stationary2,wamba2016exact,al2019handbook,liu2021exactly} there exists so far no unified theory linking their solution spaces.

In this work, we identify a large family of conformal dualities for the one-dimensional time-independent cubic-quintic NLSE. 
These dualities allow us to establish conformal maps between different solutions of the cubic-quintic NLSE and even to conformally reduce the cubic-quintic to the cubic NLSE and the linear Schrödinger equation, highlighting that the lower-order equations essentially are conformal limiting cases of the cubic-quintic NLSE.
Conformal dualities are of particular interest in physics~\cite{Burkhardt_1992,nussinov2015all, ares2016mobius} with famous instances being the Kramers–Wannier duality in statistical mechanics~\cite{Kramers_1941} and the Montonen–Olive duality in quantum field theory~\cite{Montonen_1977}. 
Further, for some nonlocal NLSEs a special conformal symmetry of the time coordinate has been discussed in a different context~\cite{jackiw1990dynamical, ghosh2001conformal}.

The novel insights into the conformal duality of NLSEs presented in this Letter can directly be applied to the determination of physical parameters from noisy experimental data. 
By conformally transforming the density distributions in a nonlinear way we avoid fitting routines to getting trapped in local minima and barren plateaus~\cite{mcclean2018barren,larocca2024review} to find the global best parameters more reliably. 
This approach is of immediate interest for the study of
one-dimensional Bose-Einstein condensates~\cite{Schreck_2001,Goerlitz_2001,Greiner_2001,Meyrath_2005} described by higher-order Gross-Pitaevskii equations~\cite{Kolomeisky_2000,Abdullaev_2001,Salasnich_2002,bulgac2002dilute,Cardoso_2011,petrov2016ultradilute}
featuring two-and three-body contact interactions, and offers great potential usage also for other physical systems like for instance in nonlinear fiber optics.

\textit{Nonlinear Schrödinger equation --}
We consider the dimensionless local and time-independent cubic-quintic NLSE
\begin{align} \label{eq:NLSE}
\left(  - \frac{1}{2} \frac{\Mdiff^{2}}{\Mdiff x^{2}} + a_{3} \abs{\psi  }^{2} + \frac{a_{4}}{2} \abs{\psi  }^{4}\right) \psi  = a_2 \psi  
\end{align}
in one spatial dimension of coordinate $x$, where $\psi = \psi \left(x \right) $ is the complex-valued wave function and $a_2$, $a_3$, and $a_4$ are constants~\footnote{The indices $j$ of the coefficients $a_j$ are chosen such that they correspond to the exponents of the different powers of $\sigma$ of the polynomial P defined in Eq.~\eqref{eq:Polynom}}.
By omitting position-dependent potentials, we focus on the homogeneous case with either box or periodic boundary conditions. 

The amplitude-phase representation 
$\psi \equiv \sqrt{\sigma} \exp \left( i \phi \right)$
casts Eq.~\eqref{eq:NLSE} into the differential equations~\cite{gagnon1989exact,schurmann1996traveling,crosta2011bistability} 
\begin{align} \label{eq:DGL:density}
\left( \frac{\Mdiff \sigma }{\Mdiff x} \right)^2 = P \left(\sigma \right) 
\end{align}
for the density $\sigma = \sigma(x)$ with the quartic polynomial
\begin{align}\label{eq:Polynom}
  P \left(\sigma \right)  \equiv \frac{4}{3} a_4 \,  \sigma^4 + 4a_3 \, \sigma^3 - 8 a_2  \, \sigma^2 -16 a_1\, \sigma -4 a_0 
\end{align}
and 
\begin{align}
\frac{\Mdiff \phi }{\Mdiff x} &= \pm\frac{\sqrt{a_0}}{\sigma } 
\label{eq:DGL:Phase} 
\end{align}
for the phase $\phi = \phi(x)$. Here $a_0, a_1 $ are constants of integration and the different signs in Eq.~\eqref{eq:DGL:Phase} refer to the two possible directions of the flow induced by the phase gradient.

Obviously, the order of the polynomial $P = P(\sigma)$ directly depends on the leading nonlinearity in Eq.~\eqref{eq:NLSE} yielding a cubic or quartic polynomial in the case of the cubic ($a_4 = 0$) or cubic-quintic NLSE, while the polynomial is quadratic for the linear Schrödinger equation ($a_3 = a_4 = 0$).

\textit{Classification of solutions --}
The stationary solutions of Eq.~\eqref{eq:NLSE} are in general determined~\cite{schurmann1996traveling} by the polynomial $P$, defined by Eq.~\eqref{eq:Polynom}, and its discriminant $\Delta = a_4^6 \prod_{j \neq k} (\sigma_j - \sigma_k)$ given by the roots $\sigma_j$ of $P$.  
Depending on the sign of $\Delta$, three classes of solutions can be identified: (i) simple complex conjugated roots ($\Delta < 0$), (ii) multiple roots ($\Delta = 0$), or (iii) only simple roots ($\Delta > 0$) of $P$.

\begin{figure}[h]
\includegraphics[width=0.45\textwidth]{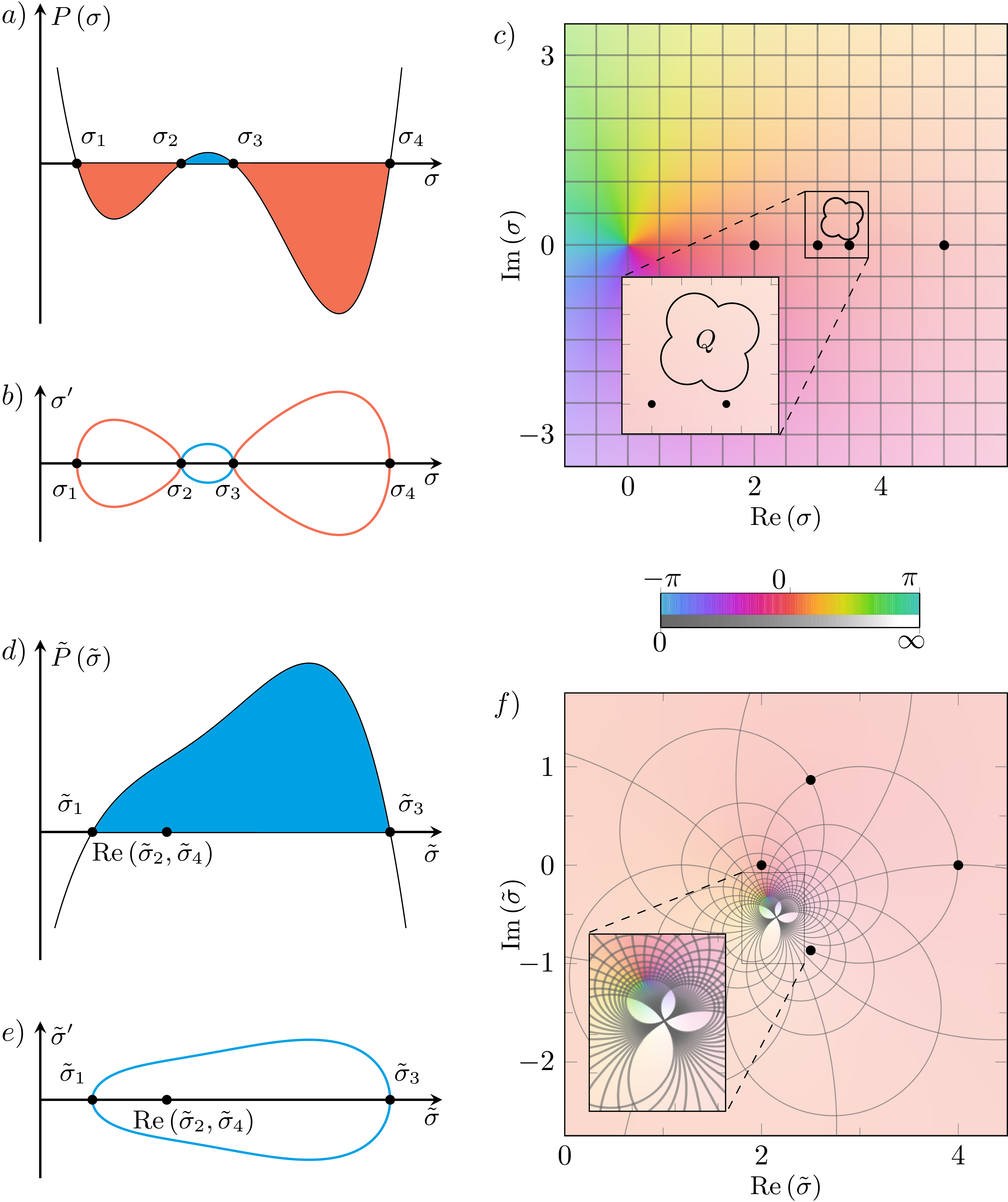}
 \caption{Conformal mapping between two realizations of the cubic-quintic NLSE with discriminant $\Delta > 0$ (a--c) and $\Delta < 0$ (d--f).
 (a,d) Polynomials $P(\sigma)$ and $\tilde{P}(\tilde\sigma)$, defined by Eq.~\eqref{eq:DGL:density}, with four simple roots $\sigma_j$ (a) or two simple and two complex roots $\tilde\sigma_j$ (d), respectively. 
 (b,e) Oscillating phase-space trajectories corresponding to real (blue) and complex (red) solutions determined by $P$  in (a,d).
 The two cases (a--c) and (d--f) are related by a conformal transformation, Eq.~\eqref{eq:ConformalMap}, which maps the positions of the roots, the polynomials, and the corresponding phase-space orbits into each other. 
 (c,f) Complex density plane of $P$ and $\tilde{P}$ shown in (a,d) illustrating the positions of the roots $\sigma_j$ and $\tilde\sigma_j$ (black dots) as well as the argument of the phase of the density $\sigma$ (color map). 
 The lines of constant real and imaginary part of the density $\sigma$ form a square grid (c) which is mapped into a grid of circles (f) by the conformal transformation due to changing the sign of $\Delta$. The roots $\sigma_j$ are thus mapped from a straight line (c) to a circle (f) with counter-clockwise orientation starting from the first real root. 
 Likewise, the cloverleaf-shaped boundary $Q$ shown in (c) is the inverse image of the square boundary shown in (f) while the center of the angle-shaped region in (f) corresponds to the point at infinity in (c).}
 \label{fig:1}
\end{figure}

In order to discuss the roots $\sigma_j$ of $P$ it is convenient to introduce the tuple notation ($r_4$, $r_3$, $r_2$, $r_1$), where every entry $r_m$ denotes the number of roots at order $m$. For instance $(0,0,0,4)$ labels a polynomial with four simple real roots as displayed in Fig.~\ref{fig:1}a, while a polynomial with two simple real roots and two simple complex-conjugated roots as shown in Fig.~\ref{fig:1}d is labeled by $(0,0,0,2+2_{{\mathbb{C}}})$.

The explicit solutions of the stationary NLSE are obtained by direct integration of Eqs.~\eqref{eq:DGL:density} and \eqref{eq:DGL:Phase} in the region between two neighboring real roots.
Consequently, oscillatory solutions of Eq.~\eqref{eq:DGL:density} occur between two neighboring simple real roots which define the minimum and maximum density of the oscillation as displayed by the three closed phase-space orbits in Fig.~\ref{fig:1}b which originate from the polynomial in Fig.~\ref{fig:1}a.

Complex conjugate roots with finite imaginary parts can therefore not be the turning points of such solutions, but instead deform the resulting orbits spanned between other real roots as illustrated in Fig.~\ref{fig:1}e. Due to the finite order of $P$, there is only one oscillatory orbit possible for the polynomial shown in Fig.~\ref{fig:1}d.

\begin{figure}
\begin{center}
\includegraphics[width=0.48\textwidth]{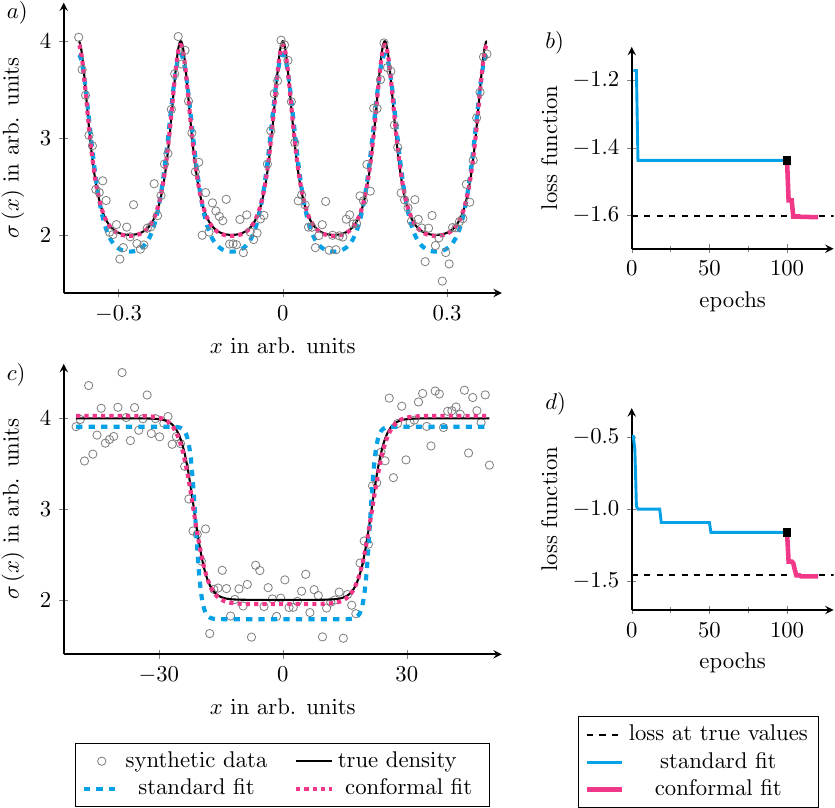}
 \end{center}
 \caption{
 Transformation-enhanced fitting of oscillating (a,b) and solitonic (c,d) solutions of the cubic-quintic NLSE. 
 (a,c) Synthetically generated noisy densities $\sigma(x)$ and the results of fitting them directly (blue) or in conformal space (red) indicating the improved accuracy of the conformal fit with respect to the original density (black).
 (b,d) Minimal loss function value as a function of fitting the data over several epochs. The standard fit (blue) applies a new random initial guess for the fit parameters every epoch, but does not improve further after 10s of epochs. Starting from 100 epochs (black square) the process is enhanced by an afterburner utilizing random conformal maps, Eq.~\eqref{eq:ConformalMap}, of the density and fitting the transformed data in the conformal space (red) yielding better loss function values even undercutting the value of the true parameters (black dashed line) after few additional epochs.}
 \label{fig:3}
\end{figure}

For polynomials with a multiple root (shown in Fig.~\ref{fig:2}b for the case (0,0,1,2)), solitonic and other more exotic solutions emerge~\footnote{See Supplemental Material for details on the conformal transformation of the NLSE, explicit expressions of the solutions discussed in Fig.~\ref{fig:2}, the connection of the conformal duality to Newtonian mechanics as well as details on the conformal fitting procedure and explicit results of the parameter estimation shown in Fig.~\ref{fig:3}. The Supplemental Material cites Refs.~\cite{bateman1953higher, milne1911elementary, dauxois2006physics, crosta2011bistability, byrd2013handbook}.}. 
In fact, the multiple root acts as a bifurcation point for the phase-space trajectories separating the two other solution classes~\cite{crosta2011bistability}. Moreover, there always exists a constant amplitude solution at the density value of the multiple root. 

Finally, the outer density regions which are restricted by only one real root typically lead to unbounded solutions. For instance, the light orange shaded region of the polynomial displayed in Fig.~\ref{fig:2}c yields such an unbounded solution~\cite{Note2}.

Consequently, the sign of the discriminant $\Delta$ and thus the nature of the roots of $P$ not only determine the character and shape of the resulting solutions, but also the total number of different solutions for a given set of parameters. 
Indeed, according to Eq.~\eqref{eq:DGL:density} physically meaningful real solutions require $P(\sigma) > 0$ between the roots considered, in addition to any restrictions set by the boundary conditions of the system under study, while for $P(\sigma) < 0$ complex density solutions emerge. 
Hence, this approach enables a straightforward and systematic classification of all possible stationary solutions of higher-order NLSEs.

\textit{Conformal duality --}
In the phase space $(\sigma, \sigma')$ with $\sigma^{\prime} \equiv \Mdiff \sigma/\Mdiff x$, the differential equation Eq.~\eqref{eq:DGL:density} constitutes an elliptic curve. A key characteristic of elliptic curves is the possibility to transform their underlying algebraic equation by rational transformations~\cite{bateman1953higher}.
In the case of the NLSE the Möbius transformation can be adapted to the differential equation Eq.~\eqref{eq:DGL:density} 
leading to the conformal map
\begin{align} \label{eq:ConformalMap}
\sigma \left( x \right) = \frac{A \, \tilde{\sigma} \left( \tilde{x}\right) + B}{C \, \tilde{\sigma} \left( \tilde{x} \right) +D} 
\end{align}
of the densities $\sigma$ and $\tilde\sigma$ with the generally complex-valued coefficients $A,B,C,D $.
In contrast to the mapping of elliptic curves, here the spatial coordinate $x$ also needs to be transformed according to the affine transformation $x = x_0 + \left(AD-BC \right) \tilde{x}$, where $x$ can become complex-valued and $x_0$ is a constant. 

The duality, Eq.~\eqref{eq:ConformalMap}, relates any two physical systems with the same real-valued cross-ratio $k^2 \in (0,1)$ which is an invariant of the transformation determined by the roots $\sigma_j$ of $P$~\cite{Note2}. 
Note that the conformal character of the Möbius transformation will preserve angles in the complex density plane by mapping every straight line of constant density into another line or circle of constant density, and vice versa.

The gradient of the phase, determined by Eq.~\eqref{eq:DGL:Phase}, enjoys a similar transformation~\cite{Note2} 
\begin{align} \label{eq:conformalMap_Phase}
    \frac{\Mdiff \phi}{\Mdiff x} = \pm \sqrt{a_0}  \,  \dfrac{ D \, \frac{\Mdiff \tilde{\phi}}{\Mdiff \tilde{x}}  \pm \sqrt{\tilde{a}_0} \, C	} { B \, \frac{\Mdiff \tilde{\phi}}{\Mdiff \tilde{x}} \pm \sqrt{\tilde{a}_0} \, A  }
\end{align}
with the very same coefficients $A,B,C,D$ as in Eq. \eqref{eq:ConformalMap}. 
As a result, the combination of Eqs.~\eqref{eq:ConformalMap} and \eqref{eq:conformalMap_Phase} provides the complete conformal mapping of the differential equations under study. Hence, different realizations of the cubic-quintic NLSE are conformally related establishing a fundamental connection between their solution spaces. 
In particular, these transformations also apply to the solutions of density $\sigma$ and phase $\phi$ themselves such that the complete complex wave function $\psi$ can be conformally mapped. 

We emphasize that the conformal duality remains intact for traveling-wave solutions of the NLSE such as solitary waves subjected to a velocity boost. This effect is a direct consequence of the Galilean covariance of the NLSE ~\cite{Note2}.

\textit{Conformal mapping and reduction of the NLSE --}
Depending on the choice of the transformation coefficients different scenarios can be realized. 
Indeed, the conformal map, Eq.~\eqref{eq:ConformalMap}, directly relates different quartic polynomials with each other and therefore their solution spaces. In this case the ratio $A/C$ must not match the value of any of the roots as this point will be mapped to infinity. Real-valued coefficients $A,B,C,D$ connect different polynomials within a given solution class, while complex coefficients enable to change the solution class corresponding to a change of sign of $\Delta$. Figure~\ref{fig:1} shows an intriguing example of the latter case, where a $(0,0,0,4)$-polynomial is mapped to one classified by $(0,0,0,2+2_{{\mathbb{C}}})$. Despite the fact that the graphs of the polynomials $P$  and $\tilde{P}$ (a,d) and their phase-space orbits (b,e) appear quite distinct in Fig.~\ref{fig:1}, they are intimately connected as shown by the density maps (c,f). 
Here, the conformal character of the transformation manifests itself by transforming the straight line connecting all four simple roots in Fig.~\ref{fig:1}c to a circle which passes again through all (now partly complex) roots in Fig.~\ref{fig:1}f. In the same way, the rectangular boundary of the density plot in Fig.~\ref{fig:1}f is mapped into the cloverleaf-shaped boundary displayed in Fig.~\ref{fig:1}c.

\begin{figure}
\begin{center}
\includegraphics[width=0.45\textwidth]{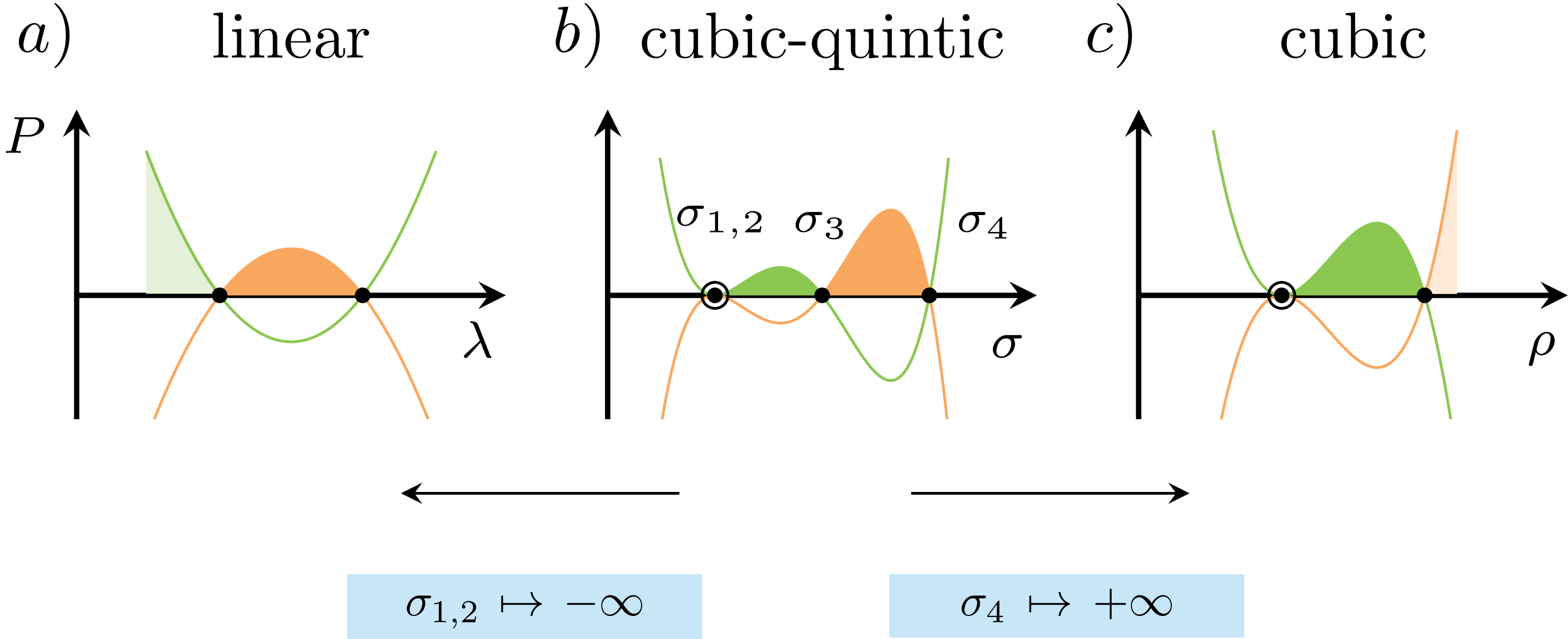}
 \end{center}
 \caption{Conformal reduction from the cubic-quintic (b) to the cubic NLSE (c) and the linear Schrödinger equation (a). 
 Exemplary polynomials $P$ (green) and $-P$ (orange), Eq.~\eqref{eq:DGL:density}, 
 with (a) two simple roots (black dots), (b) two simple roots $\sigma_3$, $\sigma_4$ and one double root $\sigma_{1,2}$ (encircled black dot), or (c) one simple and one double root.
 By moving the roots $\sigma_4$ (or $\sigma_{1,2}$) to infinity the cubic-quintic NLSE can be reduced to the cubic NLSE or the linear Schrödinger equation, respectively. 
The green (soliton solutions) and orange (oscillating solutions) fillings show which of the density regions (and corresponding solutions) are mapped into each other featuring similar characteristics. The light shaded green and orange areas in (a) and (c) illustrate unbounded solutions.}
 \label{fig:2}
\end{figure}
Moreover, as depicted in Fig.~\ref{fig:2}, it is possible to conformally reduce the cubic-quintic NLSE to either the cubic NLSE or the linear Schrödinger equation by mapping either a simple root or a double root to infinity. 
In these cases, the ratio $A/C$ must match the value of the roots to be moved. As a consequence, the overall degree of the polynomial is reduced by one (simple root moved) or two (double root moved). Analogously, the linear Schrödinger equation with an energy eigenvalue of zero is obtained by removing a triple root. Note, the cross-ratio in case of a multiple root by construction either is $k^2 = 0$ corresponding to the trigonometric limit or $k^2=1$ the hyperbolic limit~\cite{Note2}.

By reducing the degree of the polynomial, the solution space changes based on Eq.~\eqref{eq:ConformalMap} and as illustrated in Fig.~\ref{fig:2}: (i) one unbounded solution vanishes because the root constituting its minimum or maximum density has been removed, (ii) a bound solution becomes unbound since it is now only restricted by one root, and (iii) the remaining solutions get transformed, but keep their main characteristics as their roots retain their order.

The case shown in Fig.~\ref{fig:2} highlights two prominent solitonic solutions, namely the flat-top soliton~\cite{bulgac2002dilute} (green shaded area in b) and the elementary bright soliton~\cite{PethickSmith2002} (green shaded area in c) which are both governed by a hyperbolic cosine in the denominator of their density profile. By the transformation from the cubic-quintic to the cubic NLSE only the prefactor in front of the hyperbolic cosine gets changed such that both solutions are quite similar~\cite{Note2}.

Likewise, the oscillatory solution in the cubic-quintic case (orange area in Fig.~\ref{fig:2}b) governed by a cosine in the denominator as well changes its prefactor when transformed. However, in this case the corresponding solution of the cubic case (light orange area in Fig.~\ref{fig:2}c) becomes unbound due to the now different prefactor and has thus completely changed its character by the transformation.

The solutions of the green and orange regions in Fig.~\ref{fig:2} are also interconnected by a transformation that maps a real position coordinate $x$ to a purely imaginary position $\tilde x$ changing the functional dependency from a hyperbolic sine (green) to a trigonometric sine (orange). Effectively, this transformation thus flips the overall sign of the polynomial from $P$ to $-P$.
In this way all the solutions of the cubic and cubic-quintic NLSE as well as the linear Schrödinger equation are fundamentally connected.

This connection of solution spaces can also be adapted to the dynamics of classical particles in Newtonian mechanics. 
In this analogy the nonlinearites directly correspond to anharmonic conservative potentials and their solutions can be transformed into each other by means of the conformal duality~\cite{Note2}.

\textit{Transformation enhanced parameter estimation --}
As discussed above, solutions of the NLSE within a given solution class can be mapped into each other by real-valued transformation coefficients without changing the form of their analytical description.
Hence, extracting the underlying physical parameters $a_0$ to $a_4$ of one solution enables reconstruction of the parameters of the conformally connected solution as long as the transformation coefficients are known~\cite{Note2}.
We exploit this fact to improve the determination of physical parameters from experimental data. 
Figure~\ref{fig:3} shows two examples of synthetically generated noisy data sets and the results of fitting them to their known analytical functions.
Even when repeating the fitting process several times with different initial conditions a given fitting routine might get stuck in local minima of the loss function landscape (blue steps) instead of finding more truthful values (black dashed line) \cite{brastein2019estimating}.
This behavior is due to the complexity of the solution space with four (solitonic solution) or five (oscillatory solution) free parameters in each case and the additional noise which can render a reliable parameter estimation of the already highly nonlinear problem quite challenging.

To overcome this problem, we apply the conformal map, Eq~\eqref{eq:ConformalMap}, with random transformation coefficients $A,B,C,D$ to the noisy data and fit the density distribution again in the conformally transformed space. 
The extracted fitting parameters can then be transformed back to the original space and are compared with the results obtained without transformation.
Figure~\ref{fig:3} clearly shows that our conformal fitting process can substantially improve the results already after a few epochs (red steps).
The conformal transformations are changing the loss function landscape such that local minima vanish or get moved while the global minimum gets more pronounced and its location is approximately kept the same. This prevents the optimization search from getting trapped in local minima and barren plateaus. In essence, we are using the conformal transformations as an optimization afterburner. To the best of our knowledge, this strategy of using group transformations to improve the performance of search algorithms has not been used before in the literature. In this case we are using the group associated with Möbius transformations. However, the general concept of applying group transformations as a search optimization afterburner should have plenty of applications across many fields.

We have tested our method for a large variety of scenarios and observed an improvement in most cases. 
In no cases did we find that the new approach performed more poorly than the original search.
Consequently, our conformal fitting afterburner is a powerful search tool for parameter estimation of the NLSE and is applicable to a large variety of physical systems.

\textit{Conclusion --}
In summary we have provided a unified picture of the NLSE by establishing a conformal duality between the solution spaces of the cubic-quintic and cubic NLSE as well as the linear Schrödinger equation. 
Our results apply to stationary and travelling-wave solutions of the NLSE and remain valid even under Galilean transformations.
We therefore expect our findings to have a wide variety of applications that include the dynamics of solitons and their dual counterparts, mode structures in nonlinear fiber optics, hydrodynamic wave-dynamics, and the interplay of two- and three-body interactions in quasi-1D Bose-Einstein condensates as utilized for atomtronics devices~\cite{Amico_2021,Amico_2022}. 
In addition, our algebraic-geometric classification scheme can straightforwardly be extended to even higher order NLSEs such as the cubic-quintic-septic NLSE~\cite{Reyna_2017} to search for new physics in the form of exotic solutions that require  strong nonlinearities of this kind. 

Finally, we have demonstrated that the conformal duality can be applied as a resource to substantially improve parameter estimation of noisy experimental data by smoothing the loss function landscape to find the true parameters more reliably.
This direct application of the conformal duality should have far reaching consequences for the treatment of nonlinear systems in future experiments and numerical optimization.

\textit{Acknowledgments --}
We thank M.A. Efremov for fruitful discussions and helpful suggestions. D.L. acknowledges financial support from the U.S. Department of Energy (DE-SC0021152, DE-SC0013365, DE-SC0023658, SciDAC-5 NUCLEI Collaboration). 
W.P.S. is grateful to Texas A$\&$M University for a Faculty Fellowship at the Hagler Institute for Advanced Study at the Texas A$\&$M University as well as to the Texas A$\&$M AgriLife Research.

\bibliography{biblio.bib}

\onecolumngrid

\clearpage

\begin{center}
	{\normalfont\large\bfseries\centering {\it Supplemental material for}\\ \vspace{0.4cm}
	Conformal duality of the nonlinear Schrödinger equation: \\ Theory and applications to parameter estimation}\\	\vspace{0.4cm}
David B. Reinhardt$^{1}$, Dean Lee$^{2}$, Wolfgang P. Schleich$^{3,4}$ and Matthias Meister$^{1}$ \\ \vspace{0.4cm}
\textit{$^{1}$German Aerospace Center (DLR), Institute of Quantum Technologies,\\ Wilhelm-Runge-Straße 10, 89081 Ulm, Germany \\
$^{2}$Facility for Rare Isotope Beams and Department of Physics and
Astronomy,\\
Michigan State University, MI 48824, USA \\
$^{3}$Institut f\"ur Quantenphysik and Center for Integrated Quantum Science and Technology (IQ$^{\rm ST}$),\\ Universit\"at Ulm, D-89069 Ulm, Germany \\
$^{4}$Hagler Institute for Advanced Study at Texas A$\&$M University, Texas A$\&$M AgriLife Research, Institute for Quantum Science and Engineering (IQSE), and Department of Physics and Astronomy, Texas A$\&$M University, College Station, Texas 77843-4242, USA} \\ \vspace{0.4cm}
(Dated: July 4, 2024) \\ \vspace{1cm}
\end{center}

\section{A. Conformal duality of the nonlinear Schrödinger equation}

In this section we derive the transformation of the gradient of the phase from the Möbius transformation of the density and further provide more insights into the transformation of the differential equations with quartic, cubic, and quadratic polynomial dependency in the density. In addition, the cross-ratio of the transformation is discussed in more detail. 
Moreover, we provide explicit expressions for the conformal reduction of typical oscillating solutions as shown in Fig.~\ref{fig:1} and of the solitonic solutions shown in Fig.~\ref{fig:2}. In this context we also present the involved analytical solutions of the NLSE.  
Finally, we give an overview of all possible polynomials of the cubic-quintic NLSE and how they can be reduced to those of the cubic NLSE and the linear Schrödinger equation. 

\subsection{A.1 Basics of elliptic curves}
A common way to define elliptic- and hyper-elliptic curves is to introduce the relation
\begin{align}
    Y^2 = P (X)
\end{align}
between the algebraic variables $X$ and $Y$ with the polynomial 
\begin{align}
    P (X) \equiv\sum_{n=0}^{N} \alpha_n X^n
\end{align}
of order $N$ with coefficients $\alpha_n$. Note that for elliptic curves $N=3,4$, while for hyper-elliptic curves $N>4$ \cite{byrd2013handbook}.

A key feature of elliptic or hyper-elliptic curves is the possibility to transform them into other elliptic or hyper-elliptic curves by rational transformations~\cite{bateman1953higher}. For instance the bi-rational transformations
\begin{align}
    X= \frac{A  \Tilde{X} +B}{C \Tilde{X}+D}
\end{align}
and
\begin{align}\label{eq:supp:trafo_Y}
     Y = \frac{ \Tilde{Y}}{\left(CX + D \right)^{\frac{N}{2}}} 
\end{align}
connect both algebraic variables $X$ and $Y$ with the corresponding new variables $\Tilde{X}$ and $\Tilde{Y}$ through the coefficients $A,B,C$ and $D$. 

\subsection{A.2 Conformal transformation of the density and the phase gradient}

In the case of the nonlinear Schrödinger equation (NLSE) discussed here we identify the variables $X$ and $Y$ with $\sigma$ and $\sigma^{\prime} \equiv \Mdiff \sigma / \Mdiff x$ such that the transformation of the density is given by the relation 
\begin{align} \label{eq:supp:sigma_transform}
\sigma \left( x \right) = \frac{A \, \tilde{\sigma} \left( \tilde{x} \right) + B}{C \, \tilde{\sigma} \left( \tilde{x} \right) +D} 
\end{align}
being of linear fractional type.

In contrast, the transformation of the derivative of the density $\sigma^{\prime}$ is of nonlinear fractional type as is the mapping between the algebraic variables $Y$ and $\Tilde{Y}$, Eq.~\eqref{eq:supp:trafo_Y}. Nevertheless, in case of the NLSE the transformation of $\sigma^{\prime}$ is self-consistently obtained by differentiation of Eq.~\eqref{eq:supp:sigma_transform}
\begin{align}
\frac{\Mdiff  \sigma \left(x  \right)}{\Mdiff x   }  &=\frac{\Mdiff \tilde{x}   }{\Mdiff x  } \frac{AD-BC}{\left( C \,\tilde{\sigma}\left(\tilde{x}   \right) +D	 \right)^{2}  } \frac{\Mdiff \tilde{\sigma} \left(\tilde{x}    \right)}{\Mdiff \tilde{x}  } \\
&=\frac{1}{\left( C \,\tilde{\sigma}\left(\tilde{x}   \right) +D	 \right)^{2}  } \frac{\Mdiff \tilde{\sigma} \left(\tilde{x}    \right)}{\Mdiff \tilde{x}  } \,, \label{eq:supp:sigma_prime_transform}
\end{align}
where we have used the linear coordinate transformation $x = x_0 + \left(AD-BC \right)  \tilde{x}$ between $x$ and $\tilde{x}$ in the second step. 

The origin of the Möbius transformation of the phase gradient 
\begin{align}\label{eq:supp:phi_gradient}
\frac{\Mdiff \phi }{\Mdiff x} &= \pm\frac{\sqrt{a_0}}{\sigma } 
\end{align}
is rooted in its reciprocal dependence on the density $\sigma$. Indeed, by inserting the conformal map of the density, Eq.~\eqref{eq:supp:sigma_transform}, into Eq.~\eqref{eq:supp:phi_gradient} and dividing both the numerator and denominator by the density $\tilde{\sigma}$ we immediately obtain the transformation 
\begin{align}\label{eq:supp:phase_gradient_transform}
 \frac{\Mdiff \phi}{\Mdiff x} = \pm \sqrt{a_0}  \,  \dfrac{ D \, \frac{\Mdiff \tilde{\phi}}{\Mdiff \tilde{x}}  \pm \sqrt{\tilde{a}_0} \, C	} { B \, \frac{\Mdiff \tilde{\phi}}{\Mdiff \tilde{x}} \pm \sqrt{\tilde{a}_0} \, A  }   \,, 
 \end{align} 
of the phase gradient, where $\sqrt{\tilde{a}_0}$ is the flux in the transformed system. Thus, the phase gradient also transforms according to a M\"obius transform with the same coefficients $A,B,C,D$ used for the density. 
 
 Consequently, the transformations Eq.~\eqref{eq:supp:sigma_transform} and Eq.~\eqref{eq:supp:phase_gradient_transform} of the density  and gradient of the phase together enable a complete conformal mapping of the complex field $\psi = \sqrt{\sigma} \,\Me^{\Mi \phi}$ of the NLSE.

\subsection{A.3 Conformal transformation of the polynomial equation}

The stationary solutions of the NLSE are governed by the differential equation~\eqref{eq:DGL:density} which features a polynom $P (\sigma)$ of fourth order on the right-hand side. In order to analyze this equation it is convenient to consider the factored form
\begin{align}
\left( \frac{\Mdiff \sigma}{\Mdiff  x} \right)^2 &= a_{4}   \, \left( \sigma - \sigma_{1} \right) \left(\sigma-\sigma_{2} \right) \left(\sigma-\sigma_{3} \ \right) \left(\sigma-\sigma_{4} \ \right) 
\end{align}
where $\sigma_j$ are the roots of $P$.

By applying the transformations Eq.~\eqref{eq:supp:sigma_transform} and \eqref{eq:supp:sigma_prime_transform} the transformed differential equation is given by
\begin{align}\label{eq:supp:polynom_equation_quartic}
\left(\frac{\Mdiff \tilde{\sigma}}{\Mdiff \tilde{x}} \right)^2  &= a_4 \Big( \left(A-\sigma_1 C\right) \sigma+B- \sigma_1D  \Big) \cdot \Big( \left(A-\sigma_2 C\right)\sigma+B- \sigma_2D \Big)\cdot \Big(\left(A-\sigma_3 C\right)\sigma+B- \sigma_3D   \Big) \cdot \Big( \left(A-\sigma_4 C\right) \sigma+B- \sigma_4D  \Big) \nonumber \\
  &= \tilde{a}_4 \, \Big(  \tilde{\sigma} - \tilde{\sigma}_1  \Big)  \Big( \tilde{\sigma}- \tilde{\sigma}_2 \Big) 
 \Big(\tilde{\sigma} -\tilde{\sigma}_3\Big)  \Big(  \tilde{\sigma} - \tilde{\sigma}_4  \Big)
\end{align}
with 
\begin{align}
    \tilde{a_4} \equiv a_4 \, C^4  \left(\frac{A}{C} -\sigma_1 \right) \left(\frac{A}{C} -\sigma_2 \right) \left(\frac{A}{C} -\sigma_3 \right) \left(\frac{A}{C} -\sigma_4 \right) \,,
\end{align}
the new roots 
\begin{align}
    \Tilde{\sigma}_j \equiv \frac{D \sigma_j-B}{-C\sigma_j +A} 
\end{align}
and  $A/C \neq \sigma_j$. 
In general the conformal character of the transformation maps straight lines to circles and vice versa such that the new and original roots will always lie on a straight line or a circle in the complex density plane. 
By choosing a specific set of coefficients $A, B, C$ and $D$ Eq.~\eqref{eq:supp:polynom_equation_quartic} can be used to relate any two solutions of the cubic quintic NLSE with the same cross-ration $k^2$ as illustrated in Fig.~\ref{fig:1}.

Moreover, the quartic polynomial can conformally be reduced to a cubic polynomial, describing the cubic NLSE, by mapping one of the roots to infinity as shown in Fig.~\ref{fig:2}. In this case $A/C$ needs to be equal to the value of the root to be mapped. If, for instance, the first root $\sigma_1$ is mapped to infinity ($A/C = \sigma_1$) the resulting differential equation can be written as
\begin{align}\label{eq:supp:polynom_equation_cubic}
\left( \frac{\Mdiff \rho}{\Mdiff \tilde{x}} \right)^2 &= \tilde{a}_3  \left(\rho - \rho_1 \right) \left(\rho - \rho_2 \right) \left(\rho - \rho_3 \right) \\
&= \tilde{a}_3  \rho^3 + \tilde{a}_2 \rho^2 + \tilde{a}_1 \rho + \tilde{a}_0 
\end{align}
with 
\begin{align}
    \tilde{a}_3 \equiv a_{4}  C^2 \left( \frac{A}{C} -  \sigma_2   \right) \left( \frac{A}{C} -  \sigma_3   \right) \left( \frac{A}{C} -  \sigma_4  \right)   \left( BC - AD \right) \,.
\end{align}
Here we have introduced the density $\rho$  the  of the cubic NLSE as to distinguish between the various cases.

In the case that the quartic polynomial, Eq.~\eqref{eq:DGL:density}, exhibits a vanishing discriminant $\Delta$ multiple roots might emerge. If one of the roots is a double root, the degree of the polynomial can be reduced by two, giving rise to a simple parabola corresponding to the linear Schrödinger equation. By choosing e.g. $A/C=\sigma_{1,2}$ we obtain a conic curve 
\begin{align}\label{eq:supp:polynom_equation_quadratic}
\left( \frac{\Mdiff \lambda}{\Mdiff \tilde{x}} \right)^2 &=e_2 \,  \left(\lambda - \lambda_1 \right) \left(\lambda - \lambda_2 \right) \\
&= e_2 \lambda^2 + e_1 \lambda + e_0 
\end{align}
describing the stationary states of the linear Schrödinger equation, where 
\begin{align}
    e_2 \equiv a_4   \left(  \frac{A}{C} -  \sigma_3   \right) \left( \frac{A}{C} -  \sigma_4  \right)   \left( BC - AD \right)^2 
\end{align}
and the density is labeled by $\lambda$.

\subsection{A.4 Details of the cross-ratio}
In the mathematics literature \cite{milne1911elementary}, the cross-ratio is sometimes also called the anharmonic ratio of four co-linear or concyclic complex numbers $z_1,z_2,z_3$ and $z_4$ 
and is typically defined as
\begin{align}\label{eq:supp:deinition_cross_ratio}
  \left(z_1,z_2;z_3,z_4 \right) = \frac{z_3-z_1} {z_3-z_2} \frac{z_4-z_2}{z_4-z_1} \equiv \Lambda \,.  
\end{align}
Actually, this definition is not unique as the value will be unaltered for pairwise interchange of all points . Furthermore, all other possibilities of partitioning the four points can eventually be related to the cross-ratio $\Lambda$ defined in Eq.~\eqref{eq:supp:deinition_cross_ratio}.

For instance, the solutions of the ``W''-shaped graph of the polynomial $P$ shown in Fig.~\ref{fig:1}a employ a cross-ratio or elliptic modulus given by the relation 
\begin{align}
k^2 &\equiv \frac{\sigma_4-\sigma_3}{\sigma_4-\sigma_2} \frac{\sigma_2-\sigma_1}{\sigma_3-\sigma_1} = \frac{\Lambda-1}{\Lambda} \,.
\end{align}

Similarly, the solutions of a ``M''-shaped graph use the relation
\begin{align}
k^{{\prime}^2} = \frac{\sigma_3-\sigma_2}{\sigma_4-\sigma_2} \frac{\sigma_4-\sigma_1}{\sigma_3-\sigma_1} = 1 - k^2 = \frac{1}{\Lambda}
\end{align}
often referred to as the complementary elliptic modulus \cite{byrd2013handbook}. 
When employing a Möbius transform with $A/C \in (\sigma_3, \sigma_4)$ the ``W''-shaped graph is mapped to a ``M''-shaped graph. The roots get cyclically permuted in this process. More precisely this map changes the largest root of the original polynomial to the smallest root of the resulting polynomial thereby assuring the equivalency of the two cross-ratio presented.
In case of a multiple root of the polynomial $P$, the cross-ratio employed in solutions given in terms of Jacobi elliptic functions, will either correspond to the trigonometric limit where $k^2=0$ or the hyperbolic limit with $k^2=1$.

\subsection{A.5 Galilean invariance of the conformal duality}

\subsubsection{Velocity boost for the time-dependent nonlinear Schrödinger equation}

Here, we discuss the Galilean covariance of the cubic-quintic NLSE by showing that any Galilean transformation leaves the underlying equation of motion invariant. We consider a solution $\psi \left(x ,t\right)$ of the time-dependent cubic-quintic NLSE
\begin{align} \label{eq:TD-CQ-NLSE}
i  \frac{\partial}{\partial t} \psi \left(x,t \right) = \left( -\frac{1 }{2} \frac{\partial^{2}}{\partial x^{2}}  + a_3 \abs{\psi \left(x,t \right) }^{2} + a_4 \abs{\psi \left(x,t \right) }^{4}\right) \psi \left(x,t \right)
\end{align}
in the rest frame  $\mathcal{F}$ described by the coordinates $x$ and $t$. Applying a boost with the dimensionless velocity $u$ introduces the moving frame $\mathcal{F}^{\prime} $ with coordinates $x^{\prime}$ and $t^{\prime}$ which are related to the rest frame via
\begin{align}
x &= x^{\prime} + u t^{\prime} \label{eq:supp:trafo_position} \\
t &= t^{\prime} \, . \label{eq:supp:trafo_time}
\end{align}
To obtain the wave function $\psi^{\prime} \left(x^{\prime} ,t^{\prime}\right)$ in the boosted frame $\mathcal{F}^{\prime} $ we consider the transformation
\begin{align} \label{eq:wf_boosted_frame}
\psi \left(x ,t\right)  = \psi^{\prime} \left(x^{\prime} ,t^{\prime}\right) \exp \left(i  u  x^{\prime} + i \frac{u^2}{2}  t^{\prime} \right)
\end{align}
of the wavefunction $\psi \left(x ,t\right)$ in the rest frame with a phase containing the boost velocity and the additional kinetic energy due to the boost in the moving frame $\mathcal{F}^{\prime} $.

Next, we will show that the Galilean transformation of the wavefunction and coordinates leaves the NLSE in Eq.~\eqref{eq:TD-CQ-NLSE} formally unchanged. For this purpose, we consider the transformations of the derivatives in Eq.~\eqref{eq:TD-CQ-NLSE} which are given by 
\begin{align}
\frac{\partial}{\partial t} = \frac{\partial}{\partial t^{\prime}} \cdot  \frac{\partial t^{\prime} }{\partial t } + \frac{\partial}{\partial x^{\prime}} \cdot  \frac{\partial x^{\prime} }{\partial t } = \frac{\partial}{\partial t^{\prime}} - u  \frac{\partial}{\partial x^{\prime}} 
\end{align}
and 
\begin{align}
\frac{\partial}{\partial x } = \frac{\partial}{\partial x^{\prime} } \frac{\partial x^{\prime}  }{\partial x } + \frac{\partial}{\partial t^{\prime} } \frac{\partial t^{\prime}  }{\partial x } = \frac{\partial}{\partial x^{\prime} }.
\end{align}
By substituting these derivatives into Eq.~\eqref{eq:TD-CQ-NLSE} and also inserting the transformation law for the wave function, Eq.~\eqref{eq:wf_boosted_frame}, we obtain the differential equation
\begin{align} \label{eq:TD_transf_NLSE}
i  & \left( \frac{\partial}{\partial t^{\prime}} - u  \frac{\partial}{\partial x^{\prime}}  \right) \psi^{\prime} \left(x^{\prime} ,t^{\prime}\right)  \exp \left( i  u  x^{\prime} + i \frac{u^2}{2 }  t^{\prime} \right) \\
&=  \left( -\frac{ 1 }{2} \frac{\partial^{2}}{\partial {x^{\prime}}^{2}} + a_3 \abs{ \psi^{\prime} \left(x^{\prime} ,t^{\prime}\right) }^{2} + a_4 \abs{ \psi^{\prime} \left(x^{\prime} ,t^{\prime}\right) }^{4}\ \right) \psi^{\prime} \left(x^{\prime} ,t^{\prime}\right)  \exp \left( i  u  x^{\prime} + i \frac{u^2}{2}  t^{\prime} \right) \nonumber
\end{align}
describing the evolution of the boosted wave function $\psi^{\prime} \left(x^{\prime} ,t^{\prime}\right)$ in the moving frame $\mathcal{F}^{\prime} $.

By performing the partial derivatives on both sides of the equation and factoring out the global phase factor $\exp \left( i u x^{\prime} + i \frac{u^2}{2 }  t^{\prime} \right)$ the differential equation in the moving frame reduces to
\begin{align}\label{eq:supp:NLSE:boosted_final}
 i \frac{\partial }{\partial t^{\prime}} \psi^{\prime} \left(x^{\prime} ,t^{\prime}\right) =  \left( -\frac{ 1 }{2}  \frac{\partial^{2} }{\partial {x^{\prime}}^{2}} + a_3 \abs{\psi^{\prime} \left(x^{\prime} ,t^{\prime}\right)}^2 + a_4 \abs{\psi^{\prime} \left(x^{\prime} ,t^{\prime}\right)}^4 \right)  \psi^{\prime} \left(x^{\prime} ,t^{\prime}\right) \,.
\end{align}
The fact that Eqs.~\eqref{eq:TD-CQ-NLSE} and \eqref{eq:supp:NLSE:boosted_final} have the same form proves the Galilean invariance of the cubic-quintic NLSE.

\subsubsection{Impact on stationary solutions}

In order to derive the impact of the velocity boost on stationary solutions, we start with the usual ansatz
\begin{align}\label{eq:supp:stationary_state_rest:frame}
   \psi\left(x ,t\right) = \psi \left(x ,0\right)  \exp \left( -i a_2 t \right)   
\end{align}
for the time-independent wave function $\psi(x,0)$ in the rest frame $\mathcal{F}$ with eigenvalue $a_2$ determined by Eq.~\eqref{eq:NLSE}.

Applying the transformation, Eq.~\eqref{eq:wf_boosted_frame}, to both the time-dependent and time-independent wave functions in Eq.~\eqref{eq:supp:stationary_state_rest:frame} yields the relation
\begin{align} \label{eq:stationary:wf_boosted_frame}
\psi^{\prime} \left(x^{\prime} ,t^{\prime}\right) = \psi^\prime\left(x^\prime ,0\right)  \exp\left(  -i \left(a_2 + \frac{u^2}{2} \right)  t^{\prime} \right)  \,,
\end{align} 
where we have also used the relation $t = t^\prime$, Eq.~\eqref{eq:supp:trafo_time}.

Inserting this result into the time-dependent NLSE in the boosted frame, Eq.~\eqref{eq:supp:NLSE:boosted_final} allows us to obtain the time-independent NLSE in the boosted frame
\begin{align}
   \left( -\frac{ 1 }{2}  \frac{\partial^{2} }{\partial {x^{\prime}}^{2}} +   a_3 \abs{\psi^{\prime} \left(x^{\prime} ,0\right)}^2  + a_4 \abs{\psi^{\prime} \left(x^{\prime} ,0\right)}^4 \right)  \psi^{\prime} \left(x^{\prime} ,0\right)&= a_2^{\prime}  \psi^{\prime} \left(x^{\prime} ,0 \right)  \,,
\end{align}
where 
\begin{align}
a_2^{\prime} \equiv a_2 + \frac{u^2}{2 } .
\end{align}
Consequently, the boost to the moving frame adds a velocity-dependent term to the eigenvalue $a_2^\prime$ of the stationary solution.  This is just the kinetic energy associated with the Galilean boost.  

In the main text, we show that any two stationary solutions with the same cross-ratio can be converted into each other using conformal transformations. This fundamental result can now be extended to traveling wave solutions.  We simply take the dual stationary solutions and perform Galilean boosts to produce the desired traveling waves.

\subsection{A.6 Explicit solutions for the conformal duality and reduction of the nonlinear Schrödinger equation}

\subsubsection{Oscillating solutions}

As a first application of the conformal duality of the NLSE we show how typical oscillating solutions of the cubic and the cubic-quintic NLSE are connected via the transformation, Eq.~\eqref{eq:supp:sigma_transform}. For this purpose we consider the real-valued solution  
\begin{align}\label{eq:suppp:oscillating_solution_cubic}
\rho \left(x_2 \right) & = \left( \rho_2 - \rho_1 \right) \operatorname{sn}^{2} \left( \kappa_2x_2  , k \right) + \rho_{1},  
\end{align}
which oscillates between the density values $\rho_1$ and $\rho_2$ corresponding to the roots of a cubic polynomial $P (\rho)$ with an ``N''-shaped graph. Here $\operatorname{sn}$ refers to the Jacobi elliptic sine function with $\kappa_2 = a_3 \sqrt{\rho_3-\rho_1}/ 2 $ and $k^2 = (\rho_2-\rho_1)/(\rho_3-\rho_1)$. 

We now demonstrate the conformal duality between this solution and another solution for a quartic polynomial $P (\sigma)$. By applying the solution of the cubic NLSE, Eq.~\eqref{eq:suppp:oscillating_solution_cubic}, to the transformation Eq.~\eqref{eq:supp:sigma_transform}, and choosing $A/C=\sigma_1<0$, to pull a root from infinity, we obtain the solution
\begin{align}\label{eq:suppp:oscillating_solution_cubic-quintic}
 \sigma \left(x_3 \right) 
&= \frac{ \sigma_2 \eta -  \sigma_1 \operatorname{sn}^{2} \left(\kappa_3 \,  x_3 , k \right) }{  \eta -  \operatorname{sn}^{2} \left(\kappa_3 \,  x_3 , k \right) },  
\end{align}
of the cubic-quintic NLSE. Here, $\eta=\frac{\sigma_3 - \sigma_1}{\sigma_3-\sigma_2}>1$ and $\kappa_3  = \kappa_2 / (AD-BC ) $ such that the solution given by Eq.~\eqref{eq:suppp:oscillating_solution_cubic-quintic} oscillates between the densities $\sigma_2$ and $\sigma_3$ corresponding to the roots of a quartic polynomial $P (\sigma)$ with a ``W''-shaped graph. Direct integration of the quartic polynomial of the cubic-quintic NLSE yields exactly the same solution as given by Eq.~\eqref{eq:suppp:oscillating_solution_cubic-quintic} proving the validity of the conformal transformation.

The mapping coefficients of this conformal reduction are given by
\begin{alignat}{2} 
A &\equiv \frac{\sigma_1}{\rho_2-\rho_1} 
\qquad && B \equiv -\sigma_2  \eta - \sigma_1 \frac{\rho_1}{\rho_2-\rho_1} \label{eq:supp:coefficients_AB_osc} \\
C &\equiv \frac{1}{\rho_2 - \rho_1}  \qquad &&
D \equiv \eta - \frac{\rho_1}{\rho_2-\rho_1}\label{eq:supp:coefficients_CD_osc} .
\end{alignat}
The coordinates are related via the linear map $x_{3} = \left(AD-BC \right)  x_{2}  + \text{const.} $, where in this example the constant is set to zero for simplicity.

\subsubsection{Solitonic solutions}

Here, we provide the explicit transformations of the conformal reduction shown in Fig.~\ref{fig:2}.
For the bright soliton solutions we naturally need a focusing or attractive nonlinearity with $\Tilde{a}_3<0$ while the flat-top soliton likewise requires a focusing cubic nonlinearity $a_3<0$ and a defocusing quintic nonlinearity $a_4>0$. 
The analytical solutions for all orange  and green shaded areas in Fig.~\ref{fig:2} given by Eqs. \eqref{eq:supp_explicit_solutions_orange} and \eqref{eq:supp_explicit_solutions_green}, sorted according to the appearance in the figure from left to right, are given by
\begin{alignat}{3}
\lambda \left( x_1\right) &= \left(\lambda_2-\lambda_1 \right) \sin^2 \left( \kappa_1 x_1 \right) +  \lambda_1 \qquad &&\sigma \left(x_{3} \right) 
= \sigma_0 \eta \frac{1}{ 1+\sqrt{1-\eta} \cos \left(2 \kappa_3 x_{3} \right) } \qquad && \rho \left(x_{2} \right)= \rho_0 \frac{1}{1 + \cos \left(2\kappa_2 x_{2} \right) } \label{eq:supp_explicit_solutions_orange} \\
\Tilde{\lambda} \left( \tilde{x}_1 \right) &= \left(\lambda_2-\lambda_1 \right) \sinh^2 \left( \kappa_1 \tilde{x}_1 \right) +  \lambda_1 \qquad   &&\Tilde{\sigma} \left(\Tilde{x}_3 \right) 
= \sigma_0 \eta \frac{1}{ 1+\sqrt{1-\eta} \operatorname{cosh} \left(2 \kappa_3 \Tilde{x}_3 \right) } \qquad &&\tilde{\rho} \left(\tilde{x}_2 \right) = \rho_0  \frac{1}{1 + \cosh \left(2 \kappa_2 \tilde{x}_{2} \right) } \,, \label{eq:supp_explicit_solutions_green}
\end{alignat}
where $ \kappa_1 \equiv \sqrt{\abs{e_2}}$, $\rho_0 \equiv 4 \Tilde{a}_2 / \Tilde{a}_3$, $\kappa_2 \equiv \sqrt{\abs{\Tilde{a}_2}}$, $\sigma_0 \equiv -3 a_3/2a_4$, $\kappa_3 \equiv \sqrt{\abs{a_2}}$, and $\eta \equiv -8 a_2 a_4 / 3 a_3^2<1$. Again $\lambda$ refers to the density of the linear Schrödinger equation, $\rho$ corresponds to the density of the cubic NLSE, and $\sigma$ gives the density of the cubic-quintic NLSE. 

Now the transformation coefficients $A,B,C,D$, Eq.~\eqref{eq:supp:sigma_transform}, of the conformal reduction from the cubic-quintic to the cubic NLSE, connecting the solutions $\sigma \left(x_{3} \right)$ and $\rho \left(x_{2} \right)$ or $\Tilde{\sigma} \left(\Tilde{x}_3 \right) $ and $\tilde{\rho} \left(\tilde{x}_2 \right)$, can be determined. For the particular set of solutions given by Eqs.~\eqref{eq:supp_explicit_solutions_orange} and \eqref{eq:supp_explicit_solutions_green} the phase is independent of position as they require a root at the origin such that the coefficient $a_0=0$ in Eq.~\eqref{eq:DGL:density}. As a consequence, the transformation coefficient $B$  needs to be zero by construction in order to conformally relate theses solutions. In addition, this transformation requires  $A D = \frac{\kappa_2}{\kappa_3} >0$ and $A/C = \sigma_0 \left(1+ \sqrt{1- \eta } \right) =\sigma_4$. 
Ultimately, the transformation coefficients are thus given by
\begin{alignat}{2} 
A &\equiv \sqrt{\frac{\kappa_2}{\kappa_3}} \sqrt{\frac{\rho_0 \sigma_0 \eta }{\sqrt{1- \eta}}} 
\qquad && B \equiv 0 \label{eq:supp:coefficients_AB} \\
C &\equiv \sqrt{\frac{\kappa_2}{\kappa_3}} \sqrt{\frac{\rho_0}{\sigma_0}} \frac{1}{1+\sqrt{1- \eta} } \sqrt{\frac{ \eta }{\sqrt{1- \eta}}}  \qquad &&
D \equiv \sqrt{\frac{\kappa_2}{\kappa_3}}  \sqrt{\frac{\sqrt{1- \eta}}{\rho_0 \sigma_0 \eta }}. \label{eq:supp:coefficients_CD}
\end{alignat}
As the roots of the polynomial $P$ and $-P$ are identical the transformation coefficients for the conformal reduction from $\sigma \left(x_{3} \right)$ to $\rho \left(x_{2} \right)$ (orange shaded areas in Fig.~\ref{fig:2}) are formally identical to those required for the reduction from $\Tilde{\sigma} \left(\Tilde{x}_3 \right) $ to $\tilde{\rho} \left(\tilde{x}_2 \right)$ (green shaded areas in Fig.~\ref{fig:2}). Hence, the coefficients given by Eqs.~\eqref{eq:supp:coefficients_AB} and \eqref{eq:supp:coefficients_CD} apply to both cases.

Finally, in order to connect the solutions of a polynomial $P$   to those of the inverse polynomial $-P$ one can for instance employ the coefficients $A=D=(1+i)/\sqrt{2},\, B=C=0$ which formally do not change the densities according to Eq.~\eqref{eq:supp:sigma_transform}, but instead yield the coordinate transformation $x = \Mi \tilde{x}$. By transforming back to a real coordinate $\tilde{x}$ the imaginary unit is absorbed in the functional dependency of the density inducing the change from a trigonometric function to a hyperbolic function.
In this way the conformal partnering of the trigonometric oscillating solution ${\sigma} \left({x}_3 \right)$ and the resting droplet solution $\Tilde{\sigma} \left(\Tilde{x}_3 \right)$ given by Eqs.~\eqref{eq:supp_explicit_solutions_orange} and \eqref{eq:supp_explicit_solutions_green} can be realized.

\subsection{A.7 List of possible tuples for the polynomial of the cubic-quintic nonlinear Schrödinger equation}

In the main part we introduce the tuple notation ($r_4$, $r_3$, $r_2$, $r_1$), where every entry $r_m$ denotes the number of roots at order $m$ to discuss the roots $\sigma_j$ of the quartic polynomial $P$ classifying the possible solutions of the cubic-quintic NLSE.
In Table~\ref{tab:supp:tuplesclassification} we detail the complete set of tuples of the cubic-quintic NLSE and the possible conformal reductions to the cubic NLSE and the linear Schrödinger equation. Note, all mappings preserve the cross ratio $k^2$. If the discriminant $\Delta \neq 0$ the cross-ratio is given by $k^2 \in (0,1)$ while for $\Delta = 0$ the cross-ratio is either $k^2 = 0$ or $k^2 = 1$.

\begin{table}
\caption{Overview of all tuples of the cubic and cubic-quintic NLSE as well as the linear Schrödinger equation and their conformal relations for real transformation coefficients $A, B, C, D$. 
The table is organized similar to Fig.~\ref{fig:2} and the tuples are grouped by the sign of the discriminant $\Delta$. By removing a single or a multiple root through conformally mapping it to infinity the cubic-quintic tuples in most cases can be mapped to a cubic NLSE tuple (right column) or a linear Schrödinger equation tuple (left column). 
Note, cubic-quintic tuples with either $\Delta = 0$ or $\Delta \neq 0$ are also conformally related if the cross-ratio $k^2$ is the same and in gernal complex coefficients are considered like detailed in the main part.}
\label{tab:supp:tuplesclassification}
\begin{tabular}{llllll}
\\ \hline\hline \vspace{-0.1cm} \\
 \hspace{0.2cm} linear & \hspace{0.2cm}  & \hspace{0.2cm}  cubic-quintic & \hspace{0.2cm}  & \hspace{0.2cm}  cubic & \hspace{0.2cm} discriminant $\Delta$  \\
 \noalign{\vspace{2mm}} \hline \noalign{\vspace{2mm}} \\
&  &  (0,0,0,4)  &  \hspace{0.2cm} $\mapsto$  \hspace{0.2cm}  & (0,0,0,3) &  \hspace{0.2cm} $\Delta>0$  \\
  \noalign{\vspace{2mm}} \hline \noalign{\vspace{2mm}} \\
 &  &   $(0,0,0,2+2_{{\mathbb{C}}})$ & \hspace{0.2cm} $\mapsto$  \hspace{0.2cm}   &$(0,0,0,1+2_{{\mathbb{C}}})$  & \hspace{0.2cm} $\Delta<0$ 
\\ 
  &  &  $(0,0,0,4_{{\mathbb{C}}})$ & \hspace{0.2cm} $\mapsto$  \hspace{0.2cm}   &$(0,0,0,1 + 2_{{\mathbb{C}}}$) & 
\\
  \noalign{\vspace{2mm}} \hline \noalign{\vspace{2mm}} \\
      (0,0,0,2) & \hspace{0.2cm} $\mapsfrom$  \hspace{0.2cm} & (0,0,1,2) &  \hspace{0.2cm} $\mapsto$  \hspace{0.2cm}  & (0,0,1,1) &   \\
      $(0,0,0,2_{{\mathbb{C}}})$ & \hspace{0.2cm} $\mapsfrom$  \hspace{0.2cm} &$(0,0,1,2_{{\mathbb{C}}})$  &  \hspace{0.2cm} \hspace{0.2cm}  &     &   \\
  (0,0,1,0) & \hspace{0.2cm} $\mapsfrom$  \hspace{0.2cm}  & (0,0,2,0)& & & \hspace{0.2cm} $\Delta=0$   \\
 & \hspace{0.2cm}\hspace{0.2cm}  & $(0,0,2_{{\mathbb{C}}},0) $ & & \\
 (0,0,0,1) & \hspace{0.2cm} $\mapsfrom$  \hspace{0.2cm} & (0,1,0,1) &  \hspace{0.2cm} $\mapsto$  \hspace{0.2cm}   & (0,1,0,0) &  \\
  (0,0,0,0) & \hspace{0.2cm} $\mapsfrom$  \hspace{0.2cm}  & (1,0,0,0) &  & \\
\vspace{-0.2cm}
\\  \hline\hline
\end{tabular}

\end{table}

\newpage

Note, mapping a single or double complex root to infinity can result in single complex roots without a corresponding complex conjugated root like the following reductions $ (0,0,0,2+2_{{\mathbb{C}}})  \mapsto  (0,0,0,1+2_{{\mathbb{C}}})$,
$(0,0,1,2_{{\mathbb{C}}})  \mapsto     (0,0,1_{{\mathbb{C}}},1_{{\mathbb{C}}})$    or  $(0,0,1_{{\mathbb{C}}},0)  \mapsfrom (0,0,2_{{\mathbb{C}}},0) $.

\clearpage

\vspace{1cm}
\section{B. Connection of the conformal duality to Newtonian mechanics}

Besides the importance of the unified theory of the NLSE, the conformal duality discussed in this Letter has also strong implications for the dynamics of classical particles subjected to anharmonic conservative potentials in Newtonian mechanics.
Indeed, it is well-known that the NLSE formally constitutes a classical Hamiltonian system for the density $\sigma$ with the Hamiltonian function~\cite{dauxois2006physics}
\begin{align} \label{eq:Hamiltonian_system}
\mathcal{H}(\sigma', \sigma) \equiv \frac{1}{2} {\sigma^{\prime\,}}^{2} + U \left( \sigma \right) 
\end{align}
with the potential $U = U\left( \sigma \right) $. 
Here, the density $\sigma$ will be analogous to the classical position, while the spatial coordinate $x$ corresponds to time in classical mechanics.

By constraining the energy of $\mathcal{H}$ to the value of $-4a_0$ and considering the potential $U\left( \sigma \right)\equiv 2 a_0 - P \left(\sigma \right) / 2$ one can recover \cite{crosta2011bistability} Eq.~\eqref{eq:DGL:density}. 
Hence, in this analogy the nonlinearites in Eq.~\eqref{eq:NLSE} directly correspond to anharmonic contributions in $U$ with the cubic or cubic-quintic NLSE giving rise to a cubic or quartic potential, respectively, while the linear Schrödinger equation yields a harmonic potential as usual. 

The conformal map, Eq.~\eqref{eq:ConformalMap}, now allows us to transform the Hamiltonian, Eq.~\eqref{eq:Hamiltonian_system}, of a classical particle and consequently its underlying equations of motions. Thus, we can map a double-well problem to another double-well problem, or carry out the conformal reduction from a quartic to a cubic, quadratic, linear or constant potential by mapping a simple, double, triple or quadruple root of the potential to infinity. 

As a result, soliton solutions, as shown in Fig.~\ref{fig:2}, in our classical mechanics analogy correspond to an oscillation with infinitely long period where the particle in phase space approaches a bifurcation point similar to a mathematical pendulum, where the angular coordinate approaches the unstable fixed point at $\pi$ radians. 

Analogously, unbounded solutions are the counterpart of scattering states of the corresponding classical potentials. 
Hence, the ideas and concepts for treating physics problems of classical particles in anharmonic potentials have a direct correspondence to those required for the NLSE. Remarkably, both systems enable the conformal mapping of their solutions within and between different solution classes.

\clearpage

\vspace{1cm}
\section{C. Transformation-enhanced conformal fitting}
In this section we discuss the two methods we use to fit different density solutions of the CQ-NLSE subjected to Gaussian noise: the standard approach and the conformal afterburner optimization. In addition, the results of the example cases displayed in Fig.~\ref{fig:3} are presented in more detail.

\subsection{C.1 Method}
Our fitting method consists of two parts: (i) the algorithm that is used to actually perform the fit yielding results for the parameters of the model function and (ii) the way the initial conditions for each fit are generated including how the density distribution is transformed in each epoch of the conformal fitting.
While for the first part we use a standard nonlinear least-square convergence criterion to search for the minimum of the loss function, the second part actually contains the novel approach that enables our conformal afterburner to improve the results of the standard fit.

\subsubsection{Convergence criterion}
Independent whether we fit the data in the original space (standard approach) or in the conformally transformed space (conformal fit), we utilize the same fitting routine and convergence criterion to obtain values for the free parameters of the solution of the CQ-NLSE.
For the one-dimensional CQ-NLSE studied here nonlinear least-square fitting has proven to yield reliable and good results for all possible scenarios. 
Hence, the goal in each epoch is to minimize the loss function 
\begin{align}\label{eq:loss_function}
    L \left( \vec{\theta} \right)= \log_{10} \left(  \frac{\operatorname {SSR} \left( \vec{\theta} \right)}{N_{\text{Data}}} \right) \,,
\end{align}
where $\operatorname {SSR}$ refers to the sum of squared residuals. As usual the sum of squared residuals is given by
\begin{align}
     \operatorname {SSR} \left( \vec{\theta} \right) = \sum _{i=1}^{N_{\text{Data}}}(y_{i}-\sigma\left( \vec{\theta}, x_{i})\right)^{2}
\end{align}
with $\left(x_i, y_i \right)$ being the i-th of total $N_{\text{Data}}$ data points from the dataset $\{ \left(x_i, y_i \right) \}$ and $\sigma\left( \vec{\theta}, x_{i})\right)$ is one of the considered solutions of the CQ-NLSE with up to five different parameters $\vec{\theta} = \left( \kappa, \sigma_1, \sigma_2, \sigma_3, \sigma_4 \right)^T$ evaluated at $x_i$. 
For each set of initial conditions, the fitting routine thus searches for a minimum of the loss function and yields values for $\vec{\theta}$ and $\operatorname {SSR}(\vec{\theta})$ which can be compared with previous epochs. 

For the search algorithm that updates the set of parameters $\vec{\theta}$ during a single epoch to find optimal values any suitable solver can be applied in general. We have tested several possible implementations and found that our method works independent of the choice of the solver.

\subsubsection{Standard fitting procedure}
In the standard approach we try to find the optimal set of parameters to a given density distribution in the original space by varying the initial guesses within a plausible range using a uniform distribution for the initial fit parameters in each epoch.

Considering different sets of initial conditions is important as 
even slight changes in the initial conditions can lead to quite different convergence results as the loss function in our case can exhibit a plethora of different minima, non-smooth behavior and saddle-points.

After each epoch the newly obtained results are compared with the best results so far and the parameters yielding the smallest loss function are recorded. This procedure is continued for a total of $M$ standard epochs.

\subsubsection{Conformal afterburner fitting procedure}
In order to verify that the solution found by the standard fitting approach is indeed optimal or if there exists yet a better solution, we apply our conformal afterburner fitting approach after a certain number of standard epochs $M$. For that purpose we take the parameter set $\vec{\theta}^{\ast} = \left(  \kappa^{\ast}, \sigma_1^{\ast}, \sigma_2^{\ast}, \sigma_3^{\ast}, \sigma_4^{\ast} \right)^{T} $ with the smallest loss function value $L \left( \vec{\theta}^{\ast} \right)$ obtained by the standard approach as an initial guess for the first afterburner epoch. 

However, the conformal fit is not performed with the original density distribution, but we first transform the density $\sigma(x)$ to the dual space by applying the conformal map defined in Eq.~\ref{eq:ConformalMap} with randomized coefficients $A_{n}, B_{n}, C_{n}, D_{n}$ in each epoch $n$.
To enable a reliable fitting of the transformed density and a better comparison of the results in different epochs, we ensure that the minimal and maximal density values $\sigma_{\text{min/max}}$ are always mapped to the standard form values $\tilde{\sigma}_{\text{min/max}}$ by the Möbius transform
\begin{align}
\tilde{\sigma}_{\text{min/max}}  = \frac{D_{n}\sigma_{\text{min/max}} - B_{n}}{-C_{n}\sigma_{\text{min/max}} + A_{n}}  \,.
\end{align}
Here we have the freedom to choose the minimal and maximal density values $\tilde{\sigma}_{\text{min/max}}$ in the dual space. 
In both cases presented in this paper we have chosen them to keep the same values as in the original problem $\sigma_{\text{min/max}}$ though other values are generally allowed.
This prescription determines two of the transformation coefficients $A_{n}, B_{n}, C_{n}, D_{n}$ while the remaining two coefficients can be chosen randomly and in our examples are continuously uniformly distributed in a range of $a$ to $b$ with $a<b$ indicated by the operator $\operatorname{U}(a,b)$.
Thus, in every afterburner epoch $n$ a set of transformation coefficients $A_{n}, B_{n}, C_{n}, D_{n}$ is generated according to the relations

\begin{align}
A_{n} & = \operatorname{U} \left(a,b \right) \\
B_{n}  & = \operatorname{U} \left(a,b \right) \\
C_{n}  &= - B_{n}  \frac{\sigma_{\text{max}} -\sigma_{\text{min}}  }{\left(\tilde{\sigma}_{\text{max}} - \tilde{\sigma}_{\text{min}}\right) \,  \sigma_{\text{min}} \sigma_{\text{max}}} - A_{n}  \frac{
    \sigma_{\text{max}}\tilde{\sigma}_{\text{min}} - \sigma_{\text{min}} \tilde{\sigma}_{\text{max}}}{\left(\tilde{\sigma}_{\text{max}} - \tilde{\sigma}_{\text{min}} \right) \sigma_{\text{min}} \sigma_{\text{max}}} \\
D_{n}  &=  B_{n}  \frac{\sigma_{\text{max}}\tilde{\sigma}_{\text{min}} - \sigma_{\text{min}} \tilde{\sigma}_{\text{max}} }{\left(\tilde{\sigma}_{\text{max}} - \tilde{\sigma}_{\text{min}}\right) \,  \sigma_{\text{min}} \sigma_{\text{max}}} + A_{n}  \, \tilde{\sigma}_{\text{min}} \frac{\sigma_{\text{max}}\tilde{\sigma}_{\text{min}} - \sigma_{\text{min}} \tilde{\sigma}_{\text{max}} }{\left(\tilde{\sigma}_{\text{max}} - \tilde{\sigma}_{\text{min}}\right) \,  \sigma_{\text{min}} \sigma_{\text{max}}} \,
\end{align}
and is applied to map the original experimental density data $\{ \left(x_i, y_i \right) \}$ to the dual space using Eq.~\ref{eq:ConformalMap}.

Subsequently, all eligible model functions $\tilde{\sigma} (\vec{ \tilde{\theta}} , \tilde{x})$ are fitted to the conformally transformed data set $\{ \left(\tilde{x_i}, \tilde{y_i} \right) \}_{n}$ by minimizing the dual-space loss function $ \tilde{L} \left( \vec{ \tilde{\theta}} \right)$ which is defined in the same way as in the original space, Eq.~\eqref{eq:loss_function}. 
Naturally, the optimized dual-space parameter set $\vec{ \tilde{\theta}}^{\ast}_{n}$ is then back-transformed to the original space with the corresponding transformation coefficients $ \left( A_{n}, B_{n}, C_{n}, D_{n} \right)$. In addition, we evaluate the loss function $L \left( \vec{\theta} \right)$ at these back-transformed values to stay truthful to the original problem at the end of each afterburner epoch and to enable a reliable comparison of the fitting results with the ones obtained from the standard approach.

While in the first afterburner epoch we take the transformed best result $\vec{ \tilde{\theta}}$ from the standard reference approach as an initial guess, in all subsequent epochs we will update the initial guess whenever there is a further descent in the original loss function $ L \left( \vec{\theta} \right)$ due to the conformal afterburner. Therefore, we stick with the original result of the reference method in case the afterburner does not lead to an improvement.
These iterative optimization of the initial guess further enhances the conformal fitting method. However, when applying the same iterative optimization of initial guesses to the standard method one will most likely get stuck in the same minimum or descend marginally further into the very same minimum within precision bounds because the randomization of the loss function landscape is missing in the standard method.

\subsection{C.2 Example fitting results}
In this section, we present the results of the fitting parameters obtained by the standard approach and the conformal afterburner method compared to the true unbiased values as presented in Fig.~\ref{fig:3}.

\subsubsection{Oscillating solution}
For the example of an oscillating solution shown in Fig.~\ref{fig:3}a we have chosen the parameters displayed in the first row of Table~\ref{tab:supp:oscill fit parameters} to create synthetic data with the density varying between $\sigma_3$ and $\sigma_4$. 
To resemble experimental data we added $11\%$ of the density amplitude as Gaussian noise for each data point.
As shown in Table~\ref{tab:supp:oscill coefficients} the underlying $(0,0,0,4)$ polynomial has a negative leading coefficient $a_4$.

\begin{table}[H]
\caption{Comparison of the fitting parameter results of the standard and conformal method with the true values in case of the oscillating solution as displayed in Fig.~\ref{fig:3}a. The loss function value $L \left( \vec{\theta}^{\ast} \right)$ at the found minimum $\vec{\theta}^{\ast}$ and the coefficient of determination $R^2$ of the fit are also listed. 
\label{tab:supp:oscill fit parameters}}
\begin{center}
\begin{tabular}{lrrrrrrrr}
\\ \hline\hline \vspace{-0.1cm} \\
 \hspace{0.1cm} & \hspace{0.1cm}  $\kappa$ & \hspace{0.1cm}  $\sigma_1$ & \hspace{0.1cm}  $\sigma_{2}$  & \hspace{0.1cm}  $\sigma_{3}$  & \hspace{0.1cm}  $\sigma_4$ & \hspace{0.1cm}  $k^2$ & \hspace{0.1cm}  $L \left(\vec{\theta}^{\ast}\right)$ & \hspace{0.1cm}  $R^2$  \vspace{0.2cm} \\
True values & 31.21 & 0.1000 &  1.950 & 2.000 & 4.000 & 0.9500  &   -1.60 & 0.943\\
Standard fit & 33.64 & 1.762 & 1.827 & 3.863 & 2.423 $\cdot 10^{6}$ & 0.9690  & -1.44 & 0.917\\
Conformal fit & 31.40 & -0.3968 & 1.937 & 1.990 & 3.961 & 0.9522 & -1.61 & 0.944  \\
\\ \vspace{-0.2cm}
\\  \hline\hline
\end{tabular}
\end{center}
\end{table}

\begin{table}[H]
\caption{Comparison of the polynomial coefficients for the noisy oscillating solution. The coefficients were derived from the true values and the results of the standard and conformal fitting method displayed in Table~\ref{tab:supp:oscill fit parameters} by using Vietas formulas. \label{tab:supp:oscill coefficients}}
\begin{center}
\begin{tabular}{lrrrrr}
\\ \hline\hline \vspace{-0.1cm} \\
 \hspace{0.2cm} & \hspace{0.2cm}  $a_4 / 10^{3} $ & \hspace{0.2cm}  $a_3/ 10^{3}$ & \hspace{0.2cm}  $a_2 / 10^{4}$  & \hspace{0.2cm}  $a_1 / 10^{4}$  & \hspace{0.2cm}  $a_0 / 10^{3}$ \vspace{0.2cm} \\
True values & -1.00 & 8.05 & -2.05 & 1.76 & -1.56  \\
Standard fit & -8.89 $\cdot 10^{-7} $& 2.15 & -1.61 & 3.68 & -26.8 \\
Conformal fit & 0.82 & 6.12 & -1.33 & 0.62  & -4.95 \\
\\ \vspace{-0.2cm}
\\  \hline\hline
\end{tabular}
\end{center}
\end{table}

\begin{figure}[H]
\begin{center}
\includegraphics[scale=0.9]{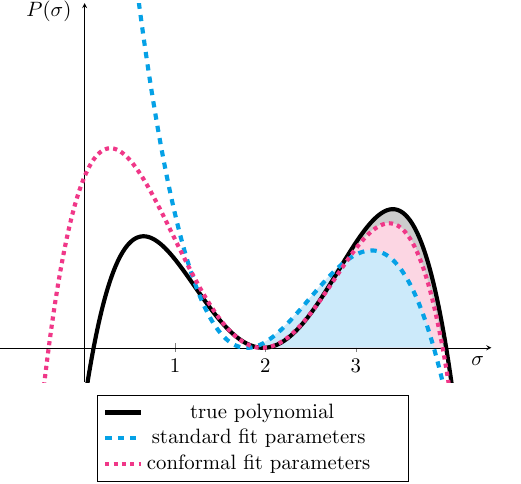}
 \end{center}
 \caption{Comparison of the quartic polynomials corresponding to the NLSE parameter estimations for the case of an oscillating solution. The coefficients are as found in Table \ref{tab:supp:oscill coefficients}. The solid black curve corresponds to the true values of the polynomial which was used to create noisy empirical data in the first place. The results of a conventional least-square minimization with bounded randomized initial guesses yields the parameter estimate corresponding to the blue dashed curve. The conformal afterburner optimizes the result of the standard method and gives a more robust parameter estimate resulting in the red dotted curve, much closer to the true polynomial. The light shaded gray, red and blue areas illustrate the considered solution between two neighboring roots of the polynomial.}
 \label{fig:supp_1}
\end{figure}
In this example the standard fitting approach gets stuck as it finds the density corresponding to  a $(0,0,0,4)$ polynomial with positive leading coefficient to be optimal, see Table~\ref{tab:supp:oscill fit parameters}. During the fit the value of the fourth root $\sigma_4$ is increased further and further by the solver until the desired tolerance is met. 
Ultimately there is no way for the solver to transition to the true polynomial employing the standard method.
This effect is observed for all tested random initial conditions.  

This fundamental problem is easily resolved by the conformal afterburner which randomly rearranges the roots but keeps the cross ratio $k^2$ constant. Since the range between $\sigma_3$ and $\sigma_4$ can get very large due to the results from the standard method, the ratio of random transformation coefficients $A/C$ will most likely fall into this interval even when considering only very few afterburner epochs. As detailed in the main part, the conformal transformation maps the point $A/C$ to infinity and everything right of this point to the left of $\sigma_1$ changing the labelling of the roots in the dual space. This effective cyclic permutation of the roots is due to the protectively extended real density line of the conformal mapping often also referred to as compactification of the real line. Therefore, $\sigma_4$ is mapped to $\tilde{\sigma}_1 + \mathcal{O} \left( 1 \right)$ making the fitting problem more accessible for the solver by changing the spacing, the order of the roots and the density model.

As a result, the conformal afterburner optimization converges to a solution very close to the true unnoisy parameters as shown in Table~\ref{tab:supp:oscill fit parameters}. 
Consequently, the physical parameters can be estimated much more accurately with the conformal fitting compared with the standard approach as illustrated by the comparison of the coefficients of the corresponding polynomial shown in Table~\ref{tab:supp:oscill coefficients} which are obtained by applying Vieta's formulas to the found fitting parameters.
In Fig.~\ref{fig:supp_1} the resulting polynomials, Eq.~\ref{eq:Polynom}, are displayed highlighting the difference between the standard approach and the conformal afterburner in comparison with the true values.

\subsubsection{Dark soliton solution}
As an example for a solitonic solution we choose the anti-flat-top dark soliton displayed in Fig.~\ref{fig:3} which is determined by a $(0,0,1,2)$ polynomial with a positive leading coefficient $a_4$. It features a minimum density given by $\sigma_2$ and a maximum density of $\sigma_{34}$. The synthetic data is created by adding 8$\%$ of the density amplitude as Gaussian noise for each data point.

\begin{table}[H]
\caption{Comparison of the fitting parameter results of the standard and conformal method with the true values in case of the dark soliton solution as shown in Fig.~\ref{fig:3}b. The loss function value $L \left( \vec{\theta}^{\ast} \right)$ at the found minimum $\vec{\theta}^{\ast}$ and the coefficient of determination $R^2$ of the fit are also listed. 
\label{tab:supp:soliton fit parameters}}
\begin{center}
\begin{tabular}{lrrrrrrr}
\\ \hline\hline \vspace{-0.1cm} \\
 \hspace{0.1cm} & \hspace{0.1cm}  $\kappa$ & \hspace{0.1cm}  $\sigma_1$ & \hspace{0.1cm}  $\sigma_{2}$  & \hspace{0.1cm}  $\sigma_{34}$ & \hspace{0.1cm} $k^2$  & \hspace{0.1cm}  $L \left( \vec{\theta}^{\ast} \right)$ & \hspace{0.1cm}  $R^2$ \vspace{0.2cm} \\
True values & 0.3163 & 1.99999 & 2 & 4 & 1 &-1.46 & 0.962 \\
Standard fit &  0.6948 & 1.787 & 1.787 &3.9067 & 1& -1.16 & 0.926 \\
Conformal fit & 0.2776 & 1.954&   1.954 & 4.027 & 1 & -1.47 & 0.963\\
\\ \vspace{-0.2cm}
\\  \hline\hline
\end{tabular}
\end{center}
\end{table}

\begin{table}[H]
\caption{Comparison of the associated polynomial coefficients for the noisy dark soliton solution. The coefficients were derived from the true values and the parameter estimates of the standard and conformal method, see Table \ref{tab:supp:soliton coefficients}, by using Vietas formulas.}
\label{tab:supp:soliton coefficients}
\begin{center}
\begin{tabular}{lrrrrr}
\\ \hline\hline \vspace{-0.1cm} \\
 \hspace{0.2cm} & \hspace{0.2cm}  $a_4$ & \hspace{0.2cm}  $a_3$ & \hspace{0.2cm}  $a_2$  & \hspace{0.2cm}  $a_1$  & \hspace{0.2cm}  $a_0$ \vspace{0.2cm} \\
True values & 0.10 & -1.20 & 5.20 & -9.60 & 6.40 \\
Standard fit & 0.43 & -4.89 & 19.9 & -34.2 & 20.9 \\
Conformal fit & 0.072 & -0.86 & 3.69 & -6.75 & 4.44 \\
\\ \vspace{-0.2cm}
\\  \hline\hline
\end{tabular}
\end{center}
\end{table}

\begin{figure}[H]
\begin{center}
\includegraphics[width=0.45\textwidth]{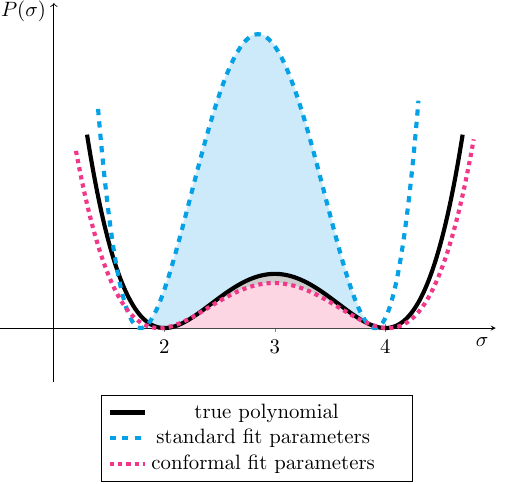}
 \end{center}
 \caption{Comparison of the quartic polynomials corresponding to the NLSE parameter estimations for the case of an dark soliton solution. The coefficients are as found in Table \ref{tab:supp:soliton coefficients}. The solid black curve corresponds to the true values of the polynomial which served as a model to create noisy empirical data in the first place. The conventional least-square minimization with bounded randomized initial guesses results in a parameter estimate corresponding to the blue dashed curve. The conformal afterburner optimizes the result of the conventional method and provides a more robust parameter estimate as can be seen from the red dotted curve. The light shaded gray, red and blue areas illustrate the considered solution between two neighboring roots of the polynomial.}
 \label{fig:supp_2}
\end{figure}
As can be seen from Table~\ref{tab:supp:soliton fit parameters} the standard fit gets stuck in a local minima, see in Fig.~\ref{fig:3} d, and fails to find the roots properly. The conformal method notably optimizes the result of the standard fit finding an estimate much closer to the true values.
The corresponding coefficients, Table~\ref{tab:supp:soliton coefficients}, of the quartic polynomials are visualized in Fig.~\ref{fig:supp_2} and are calculated from the fitting parameters with Vieta's formulas.

\subsubsection{Implementation specifics}
Since the fitting problem is noisy and highly nonlinear we use the Nelder-Mead implementation of the "scipy" minimizer. Every epoch $n$ here consists of at most 1000 internal solver iterations if the prescribed tolerance is not met before. Gradient-based solvers like the limited memory Broyden–Fletcher–Goldfarb–Shanno bounded (L-BFGS-B) algorithm do not work well in this case even when the analytical formulas for the gradients of the loss function $\vec{\nabla}_{\vec{\theta} } L \left( \vec{\theta} \right)$ are provided to the routine. This effect is due to the various local minima, non smoothness and saddle points of the loss function. For example, there always will be a plateau for the derivative of the density with respect to one of the roots as can be seen e.g. from Eq. \ref{eq:suppp:oscillating_solution_cubic-quintic}. 

Nevertheless, we have confirmed that the conformal fitting approach works independent of the specific implementation used for the solver. Even for solvers with a relatively poor performance in the original space we observed huge improvements when fitting in the conformally transformed space.

\end{document}